# BIANCHI TYPE I ANISOTROPIC UNIVERSE WITHOUT BIG SMASH DRIVEN BY LAW OF VARIATION OF HUBBLE'S PARAMETER

ANIL KUMAR YADAV

Department of Physics, Anand Engineering College, Keetham, Agra -282 007, India

E-mail: abanilyadav@yahoo.co.in, anilyadav.physics@gmail.com

A spatially homogeneous and anisotropic Bianchi Type I universe has been studied with $\omega < -1$ without Big Smash. It is demonstrated that if cosmic dark energy behaves like a fluid with equation of state $p = \omega\rho$ (p and ρ being pressure and energy density respectively) as well as generalized chaplygin gas simultaneously, Big Rip or Big Smash problem does not arise even for equation of state parameter $\omega < -1$ unlike other phantom models, here, the scale factor for Bianchi Type I universe is found regular for all time. The present model is derived by using law of variation of Hubble's parameter from Bianchi Type I space-time and also the effective role of GCG behaviour is discussed.

## 1. INTRODUCTION AND MOTIVATIONS

One of the most important properties of FRW models is, as predicated by the inflation, the flatness, which agrees with observed cosmic microwave background radiation. Even through the universe on large scale, appears homogeneous and isotropic at the present time, there is no observational data that guarantee in an epoch prior to the recombination. In the early universe the sorts of matter fields are uncertain. The existence of anisotropy at early times is a natural phenomenon to investigate, as an attempt to clarify among other things, the local anisotropies that we observe today in galaxies, cluster and super clusters so at early time it appears appropriate to suppose a geometry that is more general than just the isotropy and homogeneous FRW geometry. A Bianchi Type I model, being the straightforward generalization of the flat FRW model, is one of the simplest models of the anisotropic universe that describes a homogeneous and spatially flat universe. Unlike FRW space-time which has the same scale factor for each of the three spatial directions, Bianchi Type I space-time has a different scale factor in each direction, thereby introducing an anisotropy to the system.

Most of the well knows models of Einstein's theory and Brans-Dicke theory for FRW metric have been considered with constant deceleration parameter. In literature several authors have considered cosmological models with constant deceleration parameter (DP) [1, 2]. Recently Reddy et al [3, 4] have presented LRS Bianchi Type I models with constant DP in scalar tensor covariant theories of gravitation. Experimental probes, during last few years suggest that present universe is spatially flat as well as it is dominated by unknown form of dark energy [4, 5].

The Bianchi Type I line- element is

$$ds^2 = -dt^2 + A^2(t)dx^2 + B^2(t)dy^2 + C^2(t)dz^2, \qquad (1)$$

Theoretically accelerated expansion of universe is obtained when the cosmological model is supposed to be dominated by a fluid obeying the equation of state p= ωρ with p as isotropic pressure, ρ as energy density and $-1 \leq \omega < -1/3$.

In the recent past, it was pointed out that the current data also allowed ω < -1 [6] Rather, in refs. [7, 8, 9], it is discussed that these data favour ω < -1 being EOS parameter for phantom dark energy. Analysis of recent Ia Supernova data support ω < -1 strongly [10,11,12].

Soon after, Caldwell [7] proposed the phantom dark energy model exhibiting cosmic doomsday of the future universe, cosmologists started making efforts to avoid the problem using ω < -1 [13,14]. In the brane world scenario, Sahni and Shtanov has obtained well-behaved expansion of the future universe without Big Rip problem with ω < -1. They have shown that acceleration is a transient phenomenon in the current universe and the future universe will re-enter matter- dominated decelerated phase [15].

It is found that general relativity based phantom model encounters "sudden future singularity" leading to divergent scale factor a(t), energy density and pressure at finite time $t = t_s$. Thus the classical approach to phantom model yields big-smash problem. For the model with "sudden future singularity" Elizalde, Nojiri and Oldintov [16] argued that near $t = t_s$, curvature invariants become very strong and energy density is very high. So, quantum effects should be dominated for $|t_s = t| <$ one unit of time, like early universe. The idea is pursued in refs. [17,18,19] and shown that an escape from the Big-Smash is possible on making quantum corrections to energy density ρ and pressure p in Bianchi Type I space-time. In the framework of Robertson-Walker cosmology, Chaplygin gas (CG) is also consider as a good source of dark energy for having negative pressure, given as

$$p = -\frac{A_0}{\rho} \qquad (2)$$

with $A_0 > 0$. Moreover, it is only gas having super symmetry generalization [20, 21]. Bertolami et al [11] have found the generalized Chaplygin gas (GCG) is better fit for latest Supernova data. In the case of GCG, equation (2) looks like

$$p = -\frac{A_0}{\rho^{1/\alpha}} \tag{3}$$

where $1 \leq \alpha < \infty$.

For $\alpha = 1$ equation (3) corresponds to equation (2).

In this paper, a different prescription for GR based Future universe, dominated by the dark energy with $\omega < -1$, is proposed which is not leading to the catastrophic situations mentioned above. The scale factor, obtained here, does not possess future singularity. In the present model, it is assumed that the dark energy behaves like GCG, obeying equation (3) as well as fluid with equation of state

$$p = \omega\rho \quad \text{with} \quad \omega < -1 \tag{4}$$

Connecting equation (3) with the hydrodynamic equation

$$\dot{\rho} = -3\frac{\dot{a}}{a}(\rho + p) \tag{5}$$

and integrating, it is obtained that

$$\rho^{(1+\alpha)/\alpha}(t) = A_0 + (\rho_0^{(1+\alpha)/\alpha} - A_0)(a_0/a(t))^{3(1+\alpha)/\alpha} \tag{6}$$

with $\rho_0 = \rho(t_0)$ and $a_0 = a(t_0)$, where $t_0$ is the present time.

Equation (3 and 4) yield $\omega$ as

$$\omega(t) = -\frac{A_0}{\rho^{(1+\alpha)/\alpha}} \tag{7a}$$

So, evaluation of equation (7a) at $t = t_0$ leads to

$$A_0 = -\omega_0 \rho^{(1+\alpha)/\alpha} \tag{7b}$$

with $\omega_0 = \omega(t)$.

From equation (6) and (7), it is obtained that

$$\rho = \rho_0 \left[ -\omega_0 + (1+\omega_0)\left(\frac{a_0}{a(t)}\right)^{\frac{3(1+\alpha)}{\alpha}} \right]^{\alpha/(1+\alpha)} \tag{8}$$

with $\omega_0 < -1$.

In the homogeneous model of universe, a scalar field $\Phi(t)$ with potential $V(\Phi)$ has the energy density

$$\rho_\Phi = \frac{1}{2}\dot{\Phi}^2 + V(\Phi) \tag{9a}$$

and pressure

$$p_\Phi \quad = \quad \frac{1}{2}\dot{\Phi}^2 - V(\Phi) \tag{9b}$$

Using equations (3), (4), (7), and (8), it is obtained that

$$\dot{\Phi}^2 \quad = \quad \frac{\rho^{(1+\alpha)/\alpha} + \rho_0^{(1+\alpha)/\alpha}\omega_0}{\rho} \tag{10}$$

Connecting equations (8) and (10), it is obtained that

$$\dot{\Phi}^2 \quad = \quad \frac{(1+\omega_0)\rho_0^{(1+\alpha)/\alpha}(a_0/a)^{3(1+\alpha)/\alpha}}{[-\omega_0 + (1+\omega_0)(a_0/a)^{\frac{3(1+\alpha)}{\alpha}}]^{\alpha/(1+\alpha)}} \tag{11}$$

This equation shows that $\dot{\Phi}^2 > 0$ (giving positive kinetic energy) for $\omega_0 > -1$, which is the case of quintessence and $\dot{\Phi}^2 < 0$ (giving negative kinetic energy) for $\omega_0 < -1$, being the case of super-quintessence. As a reference, it is relevant to mention that long back, Hoyle and Narlikar used C-field (a scalar called creation) with negative kinetic energy for steady state theory of the universe [26]. Thus, it is shown that the dual behavior of dark energy fluid, obeying equation (3) and (4) is possible for scalars, frequently used for cosmological dynamics. So, this assumption is not unrealistic.

## 2. LAW OF VARIATION OF HUBBLE'S PARAMETER

Now we define, $a = (ABC)^{1/3}$ as the average scale factor so that the Hubble's parameter in anisotropic models may be defined as

$$H \quad = \quad \frac{1}{3}\left(\frac{\dot{A}}{A} + \frac{\dot{B}}{B} + \frac{\dot{C}}{C}\right) \tag{12}$$

Where an over dot denotes derivatives with respect to cosmic time t.

Also we have

$$H \quad = \quad \frac{1}{3}(H_1 + H_2 + H_3) \tag{13}$$

Where, $H_1 = \frac{\dot{A}}{A}$, $H_2 = \frac{\dot{B}}{B}$, $H_3 = \frac{\dot{C}}{C}$ are directional Hubble's factor in the direction of x, y and z-axis respectively. Very recently, S. Kumar and C. P. Singh [2] have investigated a spatially homogeneous and anisotropic Bianchi Type I model by applying a special law of variation of Hubble's parameter that yield a constant value of DP. The law of variation of Hubble's parameter is

$$H \quad = \quad Da^{-n} \quad = \quad D(ABC)^{-n} \tag{14}$$

where D and n are positive constant.

The deceleration parameter (q) is given by

$$q = -\frac{a\ddot{a}}{\dot{a}^2} \tag{15}$$

From equation (12) and (14), we get

$$\frac{\dot{a}}{a} = Da^{-n} \tag{16}$$

Integrating equation (16), we obtain

$$a = (nDt + c_0)^{1/n} \tag{17}$$

where $c_0$ is constant of integration.

Equation (17) represents accelerated expansion of universe with a(t) → ∞ as t → ∞, and support observational evidences of Ia Supernova [22,23] and WMAP [24, 25]. It is interesting to see that expansion, obtained here, is free from "finite time future singularity" unlike other General Relativity based phantom models. It is due to GCG behavior of phantom dark energy.

In this case, Hubble distance is given by

$$H^{-1} = nt + D_0 \tag{18}$$

where $D_0 = \frac{c_0}{D}$ is constant.

Equation (18) is showing the growth of Hubble's distance with time such that t → 0, $H^{-1}$ → $D_0$ and t → ∞, $H^{-1}$ → ∞. This means that, in present case galaxies will not disappear when t → ∞ unlike the phantom models with future singularity, where galaxies are expected to vanish near future singularity.

The horizon distance for this case is obtained as

$$d_H = a(t)\int_0^t \frac{dt'}{a(t')} \tag{19}$$

From equation (17) and (19), we have

$$d_H = \frac{1}{D(n-1)}[(nDt + c_0) - D'(nDt + c_0)^{\frac{1}{n}}] \tag{20}$$

Equation (20) showing that

$$d_H(t) > a(t) \text{ for } t > \frac{(1+D')^{\frac{n}{n-1}} - c_0}{nD}$$

So, horizon grows more rapidly than scale factor implying colder and darker universe. It is like flat or open universe with dominance of dark energy.

Using equation (17) and (8) the expression for energy density is given by

$$\rho = \rho_0[-\omega_0 + (1+\omega_0)(\frac{a_0}{(nDt+c_0)^{\frac{1}{n}}})^{\frac{3(1+\alpha)}{\alpha}}]^{\alpha/1+\alpha} \qquad (21)$$

## 3. DISCUSSION

The Bianchi type I anisotropic universe without big smash is driven by using the law of variation of Hubble's parameter. It is found that dark energy behaves like a fluid as well as generalized chaplygin gas simultaneously and big smash problem do not arise unlike other phantom models. Since $\omega_0 < -1$, hence from equation (21), it is clear that energy density increases with time. Also pressure increases with time. It is unlike GR based models driven by equation of state p =ωρ, where ω < -1 having future singularity at $t = t_s$, where ρ and p are divergent [7, 13] or ρ is finite and p is divergent [17, 27]. Based on Ia supernova data, Singh et al [12] have estimated $\omega_0$ for model in the range $-2.4 < \omega_0 < -1.74$ up to 95% confidence level. Taking this estimate as an example with $\alpha = 3, \rho_\infty = \rho(t \to \infty)$ is found in the range $1.51\rho_0 < \rho_\infty < 1.92\rho_0$. This does not yield much increase in ρ as t→∞ but if this model is realistic and future experiments support large $|\omega_0|$, $\rho_\infty$ will be very high. In both cases, small or large value of $\omega_0$, increases in ρ indicate creation of Phantom dark energy in future. It may be due to decay of some other components of energy in universe, which is not dominating, for example cold dark matter. It is interesting to see that Big smash problem does not arise in the present model. In refs [16, 17, 18, 19] for models with future singularity, escape from cosmic doomsday is demonstrated using quantum correction in field equations near $t = t_s$. Here using classical approach, a model for phantom cosmology, with accelerated expansion, is explored with is free from catastrophic situations. This model is derived from Bianchi type I space time using effective role of GCG behavior in natural way. Srivastav [28] have investigated FRW model with ω < -1 without Big Smash and found $1.15\rho_0 < \rho_\infty < 1.24\rho_0$ where as the present model is Bianchi Type I universe with ω < -1 without Big Smash. In this case, we found $1.51\rho_0 < \rho_\infty < 1.92\rho_0$ which is much closed to recently Ia supernova data.

*Acknowledgements:* The author would like to thanks the referee for his valuable suggestion. Also the Author is thankful to his wife Anju Yadav for her heartiest co-operation during the preparation of the manuscript.

## REFERENCES

1. S. Kumar, C.P. Singh, Astrophys. Space Sci. **312**, 57, (2007).

2. S. Kumar, C.P. Singh, Int. J. modern Phys. A, **23,** 6, 813, (2008).

3. D.R.K Reddy, M.V. Subba, G.K.Rao, Astrophys. Space Sci. **306**, 171, (2006).

4. D.R.K Reddy, R.L Naidoo,K.S.Adhav , Astrophys. Space Sci. **307**, 211, (2005).

5. A.D.Miller et al Astrophys. J. Lett., **524**, L1, (1999).

6.V.Faraoni,Phys. Rev. D **68,** 063508, (2003).

7. R.R. Caldwell, Phys. Lett. B **545**, 23, (2002).

8. H. Ziaeepour, astro-ph/0002400.

9. J. M. Cline et al, Phys. Rev.D **70**, 043543, (2004).

10.U. Alam et al, astro-ph/0311364.

11.O.Bertolami et al 2004, MNRAS,**353**, 329.

12. P. Singh, M. Sami and N.Dadhich, Phys. Rev. D, **68,** 023522,(2003).

13. B. McInnes, JHEP, **08**, 029, (2002).

14. V. K. Onemli et al, Class. Quan. Grav. **19**, 4607, (2002).

15. V. Shani Yu.V. Shtanov, JCAP, **0311**, 14, (2003)

16. E.Elizalde, S. Noriji, S.D. Odintov, Phys. Rev. D **70**, 43539, (2004).

17.S. Nojiri, S. D. Odintsov, Phys. Lett. B. **595**, 1,(2004).

18. S. K. Srivastava, JCAP **0601**, 003, (2006).

19. S. Nojiri and S. D. Odintsov and Tsujikawa, Phys. Rev. D **72**, 051703, (2005).

20. R. Jackiw, 'Lecture on supersymmetric non abelian fluid mechanics and d-branes', physics/0010042

21. M. C. Bento, O. Betrolami, A. A. Sen, Phys. Rev. D **66**, 43507, (2002).

22. S. Permutter, Astrophys. J. **517**, 565, (1999).

23. A.G. Rieses et al 1998, Astron. J. **116,** 1009.

24. D. N Spergel. et al, Astrophys. J. Suppl. **148**, 175, (2003).

25. L. Page et al, Astrophys. J. Suppl. **148**, 175, (2003).

26. F. Hoyle, J. V. Narlikar, MNRAS **108**, 372, (1948).

27. J. Barrow, Class. Quant. Grav. **21**, L82, (2004).

28. S. K. Srivastava, Phys. Lett. B **619**, 1, (2005).